\documentclass[a4paper,conference]{IEEEtran}
\usepackage{amsmath,amssymb,amsfonts}
\usepackage[table]{xcolor}
\usepackage{graphicx}
\usepackage{textcomp}
\usepackage{listings}
\usepackage{graphicx}
\usepackage{lipsum}
\usepackage{booktabs}
\usepackage{float}
\usepackage{comment}
\usepackage[tracking]{microtype}
\usepackage[utf8]{inputenc}
\usepackage{subfig}
\usepackage{cite}
\usepackage{orcidlink} 
\usepackage{tabularx}
\usepackage{makecell}
\usepackage{hyperref}
\hypersetup{hidelinks} 
\usepackage{pifont}
\usepackage{tikz}
\usetikzlibrary{arrows.meta,positioning,shapes.geometric,shapes.multipart,fit,backgrounds,calc}
\usepackage{pgfplots}
\pgfplotsset{compat=1.17}

\usepackage{eso-pic}
\newcommand{\IEEECopyrightNotice}{%
    \AddToShipoutPictureFG*{%
        \AtPageLowerLeft{%
            \put(\LenToUnit{0.65in},\LenToUnit{0.35in}){%
                \parbox{\dimexpr\paperwidth-1.3in\relax}{%
                    \fontsize{8}{9.5}\selectfont
                    \textcopyright{} 2026 IEEE. Personal use of this material is permitted. Permission from IEEE must be obtained for all other uses, in any current or future media, including reprinting/republishing this material for advertising or promotional purposes, creating new collective works, for resale or redistribution to servers or lists, or reuse of any copyrighted component of this work in other works.
                    \par
                }%
            }%
        }%
    }%
}

\begin{document}

\title{Structure-Aware Merkle Trees for Tamper-Localizable Dataset and Model Integrity on Blockchain and Distributed Storage}

\author{Zeeshan Ali\textsuperscript{* \LARGE \orcidlink{0000-0002-3577-7413}}, and Dajana Cassioli \textsuperscript{\LARGE \orcidlink{0000-0001-7574-5099}}
\\
\IEEEauthorblockA{
Department of Information Engineering, Computer Science, and Mathematics (DISIM)\\
University of L'Aquila, L'Aquila, Italy\\
}
\thanks{\textsuperscript{*}Corresponding author: Zeeshan ALI}
}

\IEEEoverridecommandlockouts
\maketitle
\IEEECopyrightNotice

\begin{abstract}
Datasets and machine learning (ML) models are often exchanged between organizations that do not fully trust each other. Their integrity is usually checked with a single hash of the whole file, which reveals \emph{that} a file was altered but not \emph{which part}, so a verifier cannot tell whether one label was flipped or an entire model layer was replaced. Existing blockchain and Merkle-tree-based integrity systems split files into rows or fixed-size chunks and share the same limitation. We propose a structure-aware Merkle tree that follows the artifact's own structure: tabular datasets are hashed column by column and \texttt{safetensors} model checkpoints tensor by tensor, with each leaf bound to its column or tensor name. Only the root of the tree and a tag identifying the hashing scheme are anchored on a public blockchain smart contract, while the file itself can be stored on a distributed storage file system (DSFS). A verifier recomputes the tree from the received file and compares its root with the anchored one; if they differ, it compares the tree with that derived from an authentic copy of the file, itself checked against the same root, and names the exact column, row, or tensor that changed, without a trusted third party. We implemented the framework as a working prototype and evaluated it on synthetic datasets and models. Every modification was detected, and every changed cell or tensor value was pinpointed exactly, whereas a conventional Merkle tree could only point to a whole row or a large block of the file. The cost is a longer, one-time hashing step for the data owner, which however remained under one second in all tests.
\end{abstract}


\IEEEpeerreviewmaketitle

\section{Introduction}
\label{sec:introduction}

Datasets and machine learning (ML) model checkpoints rarely stay within one trust domain. A dataset may be collected, curated, and consumed by different organizations, and a checkpoint trained, published on a public hub, fine-tuned, and deployed by different parties. At every hand-off, the receiver must answer a deceptively simple question: is this exactly the artifact the sender intended to send, or has it been altered, accidentally through storage corruption or maliciously? Today that question is almost always answered with a single Secure Hash Algorithm, 256-bit (SHA-256) digest computed over the entire file and compared with a previously recorded value.

A whole-file digest is cheap to compute but gives only a one-bit answer: \emph{match} or \emph{no match}. If one label in a 100{,}000-row training set is flipped by a data-poisoning attack, or one attention layer of a multi-gigabyte checkpoint is replaced by a backdoored tensor, the verifier learns only that \emph{something} differs. Finding \emph{what} and \emph{where} requires a second trustworthy copy of the whole artifact and an out-of-band diff. This gap matters in practice. Data-poisoning and backdoor attacks are well documented \cite{goldblum2020dataset,cina2022wild,tian2022comprehensive}, with demonstrated impact in healthcare diagnosis \cite{kermani2015systematic} and network intrusion detection \cite{yerlikaya2022data}, and supply-chain analyses of public model hubs find many artifacts with compliance risk or unverifiable provenance \cite{meiklejohn2025mlsupplychain,hanailscg2026}.

Recording the reference digest is a second problem, since the administrator of a centralized hash registry can silently rewrite history. Public blockchains provide an append-only, publicly auditable ledger without a controlling party \cite{nakamoto2008bitcoin,buterin2014ethereum}, and a growing body of work anchors Merkle roots of outsourced cloud data on-chain to remove the trusted auditor \cite{wang2024blockchain,han2024blockchain,liu2023data,mohan2020merkle,wei2020blockchain,he2020blockchain}. Because storing bulk data on-chain is economically infeasible, such systems keep the data in an off-chain DSFS, most commonly the content-addressed InterPlanetary File System (IPFS) \cite{benet2014ipfs}, and anchor only its content identifier (CID) and a hash \cite{kumar2020blockchain,nyaletey2019blockipfs,saif2024efficient,zhu2026research}. These systems, however, split data into fixed-size chunks and therefore report tampering as an opaque chunk index rather than as a column, row, or tensor that a data scientist can act on.

This paper combines structure-aware hashing with decentralized anchoring in a working system, i.e., the Merkle Integrity Dashboard. Instead of hashing an artifact as a byte stream, we hash it \emph{along its own structure}: one sub-tree per named column for Comma/Tab-Separated Values (CSV/TSV) datasets, and one leaf per named tensor for \texttt{safetensors} checkpoints, parsed directly from the format's JSON header. The root and a machine-readable tag naming the hashing scheme are anchored on a smart contract on a public testnet, and the file is optionally pinned to a local IPFS node that announces new content to the Distributed Hash Table (DHT) that IPFS uses for content discovery.
A verifier who shares nothing with the owner except the blockchain and IPFS, recomputes the root using the scheme declared on-chain and, on a mismatch, compares its tree with that of an authentic copy whose root matches the chain, 
to identify the exact column, row, or tensor that changed. 

Our contributions are:
\begin{itemize}
\item A \emph{structure-aware Merkle hashing scheme} whose leaves are bound to semantically meaningful units (column name and row index for datasets; tensor name, dtype, and shape for \texttt{safetensors} models), so that tampering is reported as an actionable unit rather than a byte offset.
\item \emph{Self-describing on-chain anchoring}: a smart contract that stores the root with a versioned hashing-scheme tag and the CID of the file pinned on a self-managed IPFS node, so that a verifier can re-verify a file, and obtain an authentic copy, without a trusted registry.
\item An \emph{evaluation on the production code} of owner/verifier hashing agreement, localization accuracy, and hashing overhead against a conventional flat Merkle tree, with a transparent account of the proof-of-concept (PoC) scope.
\end{itemize}

\section{Proposed Methodology}
\label{sec: proposedmethodology}

\subsection{System Overview and Roles}
\label{sec:system-overview}
The system defines two roles. The \emph{Data Owner} holds the authoritative copy of a dataset or model, computes its Merkle root client-side, optionally pins the file to IPFS, and anchors the root on-chain. The \emph{Verifier} holds a public record identifier and either an independently obtained copy of the file or its CID; it recomputes the root to confirm that it matches the anchored value or, if it does not, to localize where it diverges.

The owner needs a wallet to submit a transaction to the public contract, and the verifier needs read access to that contract and to the public IPFS network (or a locally supplied file).

\subsection{Threat Model and Publisher Identity}
\label{sec:threat}
We consider three parties: the data owner, who publishes an artifact; the verifier, who receives it; and the infrastructure in between, i.e., the DSFS (IPFS nodes and gateways), any mirror or partner that forwards the file, and the blockchain. Our trust assumptions are:
\begin{itemize}
\item \emph{Trusted}: the blockchain executes the contract correctly, and a record cannot be changed once written; SHA-256 is collision resistant; the owner's wallet key is not compromised.
\item \emph{Untrusted}: everything that stores or carries the file, i.e., IPFS nodes and gateways, mirrors, and the network, may be unavailable or slow, deliver corrupted or truncated bytes, or deliberately modify the file.
\end{itemize}

The attacker we consider is any party that modifies an artifact after its root has been anchored, for example by flipping labels in a dataset or replacing a model layer. Against this attacker, the system provides two guarantees. First, \emph{detection}: any modification changes the recomputed root, so the verifier never accepts a file whose root differs from the on-chain value. Second, \emph{localization}: comparing the tree with that of an authentic copy identifies the column, row, or tensor that changed, and a copy is accepted as authentic only if it reproduces the on-chain root, so a forged reference is detected as well.

The main use case of localization is forensic. A verifier that only needs a correct file can simply download the authentic copy; localization instead answers what was altered in the copy it has already received and possibly used, and hence how serious the alteration is. For example, if a model was trained on a partner's dataset whose root later fails to match the chain, downloading the authentic copy fixes future training but not the deployed model; learning that only the labels of rows 412 and 890 were flipped, rather than thousands, decides between a minor retraining and an incident response.

A remaining question is \emph{who} the owner is. The contract is permissionless: any address can anchor any root, so a record proves only that \emph{some} wallet registered a root at a given time, not that this wallet belongs to the organization claiming the artifact. A verifier therefore accepts only records submitted by an owner address that it has learned through a separate trusted channel, such as the organization's website or an identity certificate, and retrieves that owner's records through the contract's per-owner index. We chose a public blockchain over a signed transparency log~\cite{laurie2013rfc6962} because it provides an append-only, ordered, and timestamped record with no log operator to run or trust, at the cost of transaction fees; the hashing scheme itself is independent of the anchoring layer.

\subsection{Structure-Aware and Flat Merkle Hashing}
\label{sec:hashing}
A Merkle tree \cite{merkle1988digital} summarizes a set of data blocks into a single root by recursively hashing pairs of nodes: with $H$ denoting SHA-256, each leaf is $H_\ell(d) = H(\text{``leaf:''} \| d)$ for leaf content $d$, each internal node is $H(\text{``node:''} \| \text{left} \| \text{right})$, and an odd node at any level is paired with itself. A change to any block changes its leaf, every node on the path above it, and the root, so comparing two trees level by level reveals which leaves differ. Hashing runs in the browser through the native Web Crypto API. Before hashing, each value is prefixed with a short label stating what it is (\texttt{"leaf:"}, \texttt{"node:"}, \texttt{"row:"}, \texttt{"col:"} for a cell, \texttt{"column:"} for a column root, or \texttt{"tensor:"}). Two different kinds of node therefore never produce the same hash input, so an attacker cannot pass off an internal node as a data leaf, or one kind of leaf as another, to forge a valid-looking tree. This is a standard safeguard in Merkle tree implementations.

The framework considers four hashing schemes. The scheme is chosen automatically from the file type, as listed below, and recorded on-chain as a \texttt{HashScheme} tag (\texttt{flat} or \texttt{structured}), so a verifier never has to guess which one produced a root. The two \emph{structured} schemes are the contribution of this paper. The two \emph{flat} schemes are the conventional way of building a Merkle tree over a file; the system uses them only for files that the structured schemes cannot parse, and 
the proposed evaluation uses them as baselines.

\begin{figure}[t]
\centering
\begin{tikzpicture}[
  leaf/.style={draw, rounded corners, fill=blue!8, font=\tiny, minimum width=9mm, minimum height=6mm, align=center},
  node2/.style={draw, circle, fill=orange!15, font=\tiny, minimum size=6mm, inner sep=0pt},
  top/.style={draw, circle, fill=green!20, font=\scriptsize, minimum size=7mm, inner sep=0pt},
  colname/.style={font=\tiny\itshape},
  colleaf/.style={draw, rounded corners, fill=violet!10, font=\tiny, minimum width=11mm, minimum height=6mm, align=center, inner sep=1pt},
  arr/.style={-{Latex[length=1.6mm]}, thin}
]
\node[leaf] (a1) at (0,0) {col:\\row 0};
\node[leaf] (a2) at (0.95,0) {col:\\row 1};
\node[node2] (an) at (0.475,0.9) {$R_A$};
\node[colleaf] (ak) at (0.475,1.75) {column:\\amount};
\draw[arr] (a1) -- (an); \draw[arr] (a2) -- (an); \draw[arr] (an) -- (ak);

\node[leaf] (b1) at (2.3,0) {col:\\row 0};
\node[leaf] (b2) at (3.25,0) {col:\\row 1};
\node[node2] (bn) at (2.775,0.9) {$R_B$};
\node[colleaf] (bk) at (2.775,1.75) {column:\\date};
\draw[arr] (b1) -- (bn); \draw[arr] (b2) -- (bn); \draw[arr] (bn) -- (bk);

\node[leaf] (c1) at (4.6,0) {col:\\row 0};
\node[leaf] (c2) at (5.55,0) {col:\\row 1};
\node[node2] (cn) at (5.075,0.9) {$R_C$};
\node[colleaf] (ck) at (5.075,1.75) {column:\\label};
\draw[arr] (c1) -- (cn); \draw[arr] (c2) -- (cn); \draw[arr] (cn) -- (ck);

\node[top] (t1) at (1.6,2.75) {};
\node[top] (t2) at (5.075,2.75) {};
\node[top, fill=green!35] (root) at (3.3,3.7) {Root};
\draw[arr] (ak) -- (t1); \draw[arr] (bk) -- (t1);
\draw[arr] (ck) -- (t2);
\draw[arr] (t1) -- (root); \draw[arr] (t2) -- (root);
\end{tikzpicture}
\caption{Structure-aware (column-wise) Merkle tree for a tabular dataset. Each cell is a \texttt{col:} leaf bound to its column name and row index, and each column forms an independently rooted sub-tree ($R_A$, $R_B$, $R_C$). Each column root is then bound to its column name in a \texttt{column:} leaf, and the top-level root anchored on-chain is built over these leaves.}
\label{fig:merkletree}
\end{figure}
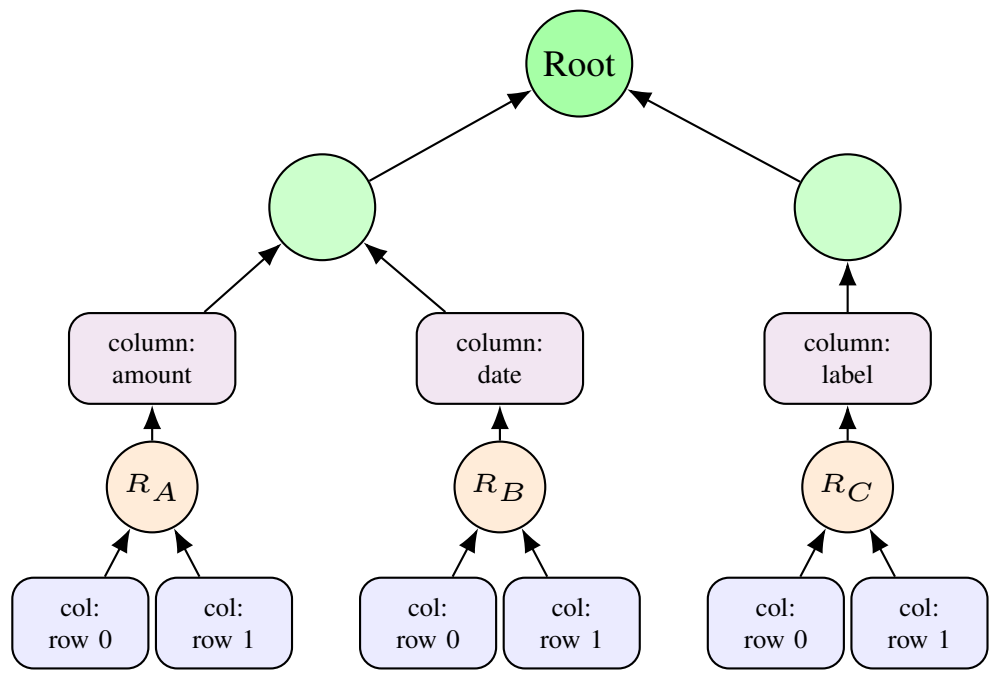

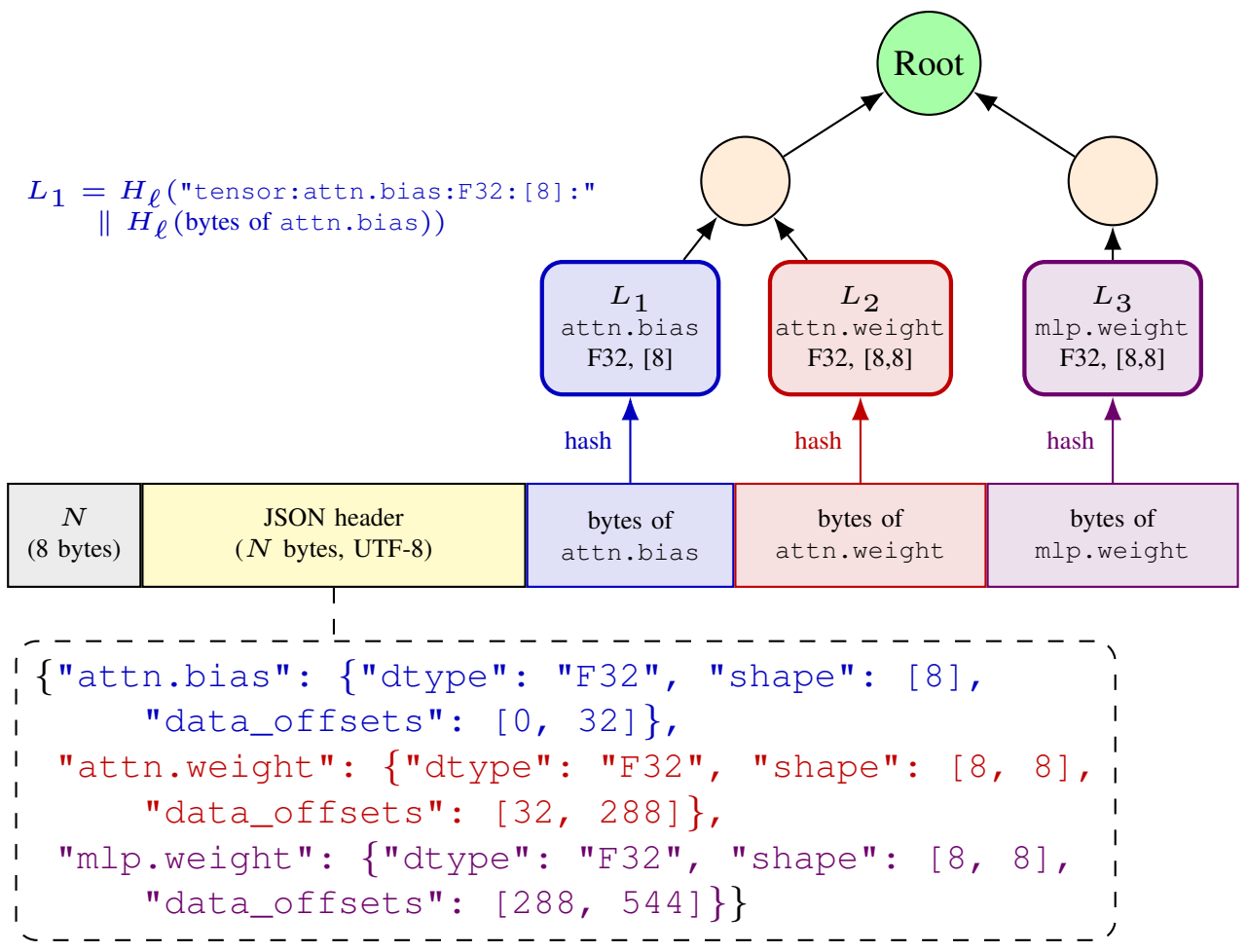
\begin{figure}[t]
\centering
\begin{tikzpicture}[
  seg/.style={draw, minimum height=7mm, font=\tiny, align=center, inner sep=1pt, anchor=south west},
  leaf/.style={draw, thick, rounded corners, font=\tiny, minimum width=12mm, minimum height=9mm, align=center, inner sep=1pt},
  node2/.style={draw, circle, fill=orange!15, font=\tiny, minimum size=6mm, inner sep=0pt},
  top/.style={draw, circle, fill=green!35, font=\scriptsize, minimum size=7mm, inner sep=0pt},
  arr/.style={-{Latex[length=1.6mm]}, thin},
  js/.style={font=\ttfamily\scriptsize, anchor=north west, inner sep=0.7pt}
]
\colorlet{tA}{blue!75!black}\colorlet{tB}{red!75!black}\colorlet{tC}{violet!85!black}

\node[seg, fill=gray!15, minimum width=9mm] (len) at (0,0) {$N$\\(8 bytes)};
\node[seg, fill=yellow!25, minimum width=26mm] (hdr) at (len.south east) {JSON header\\($N$ bytes, UTF-8)};
\node[seg, fill=tA!12, draw=tA, minimum width=14mm] (d1) at (hdr.south east) {bytes of\\\texttt{attn.bias}};
\node[seg, fill=tB!12, draw=tB, minimum width=17mm] (d2) at (d1.south east) {bytes of\\\texttt{attn.weight}};
\node[seg, fill=tC!12, draw=tC, minimum width=17mm] (d3) at (d2.south east) {bytes of\\\texttt{mlp.weight}};

\node[js] (j1) at (0.15,-0.45) {\{\textcolor{tA}{"attn.bias": \{"dtype": "F32", "shape": [8],}};
\node[js] (j2) at (j1.south west) {\hspace*{3em}\textcolor{tA}{"data\_offsets": [0, 32]\},}};
\node[js] (j3) at (j2.south west) {\ \textcolor{tB}{"attn.weight": \{"dtype": "F32", "shape": [8, 8],}};
\node[js] (j4) at (j3.south west) {\hspace*{3em}\textcolor{tB}{"data\_offsets": [32, 288]\},}};
\node[js] (j5) at (j4.south west) {\ \textcolor{tC}{"mlp.weight": \{"dtype": "F32", "shape": [8, 8],}};
\node[js] (j6) at (j5.south west) {\hspace*{3em}\textcolor{tC}{"data\_offsets": [288, 544]\}}\}};
\node[draw, dashed, rounded corners, fit=(j1)(j3)(j6), inner sep=2.5pt] (jbox) {};
\draw[dashed] (hdr.south) -- (hdr.south |- jbox.north);

\node[leaf, draw=tA, fill=tA!12] (l1) at ($(d1.north)+(0,1.05)$) {$L_1$\\\texttt{attn.bias}\\F32, [8]};
\node[leaf, draw=tB, fill=tB!12] (l2) at ($(d2.north)+(0,1.05)$) {$L_2$\\\texttt{attn.weight}\\F32, [8,8]};
\node[leaf, draw=tC, fill=tC!12] (l3) at ($(d3.north)+(0,1.05)$) {$L_3$\\\texttt{mlp.weight}\\F32, [8,8]};
\draw[arr, draw=tA] (d1.north) -- node[left, font=\tiny, text=tA]{hash} (l1.south);
\draw[arr, draw=tB] (d2.north) -- node[left, font=\tiny, text=tB]{hash} (l2.south);
\draw[arr, draw=tC] (d3.north) -- node[left, font=\tiny, text=tC]{hash} (l3.south);

\node[node2] (t1) at ($(l1)!0.5!(l2)+(0,1.0)$) {};
\node[node2] (t2) at ($(l3)+(0,1.0)$) {};
\node[top] (root) at ($(t1)!0.5!(t2)+(0,0.8)$) {Root};
\draw[arr] (l1) -- (t1); \draw[arr] (l2) -- (t1); \draw[arr] (l3) -- (t2);
\draw[arr] (t1) -- (root); \draw[arr] (t2) -- (root);

\node[font=\tiny, align=left, anchor=north west, text=tA] at (0,2.9) {$L_1 = H_\ell($\texttt{"tensor:attn.bias:F32:[8]:"}\\$\qquad\ \ \|\; H_\ell(\text{bytes of \texttt{attn.bias}}))$};
\end{tikzpicture}
\caption{Layout of a \texttt{safetensors} file and its tensor-wise Merkle tree, for the three-tensor example of Section~\ref{sec:hashing}. The first 8 bytes give the header size $N$; the JSON header lists each tensor's \texttt{dtype}, \texttt{shape}, and byte range (\texttt{data\_offsets}). Each colour marks one tensor: its leaf is built only from its own header entry (name, dtype, shape) and the hash of its own bytes, as shown for $L_1$.}
\label{fig:tensortree}
\end{figure}

\begin{itemize}
\item \emph{Structured, column-wise} is used for every CSV/TSV dataset. Each column is its own sub-tree, as shown in Figure~\ref{fig:merkletree}. Every cell becomes one leaf, obtained by hashing the cell value together with its column name and row number; for example, the value \texttt{42} in row 3 of column \texttt{amount} is hashed as \texttt{"col:amount:3:42"} (rows are counted from 0, excluding the header). The leaves of one column form a sub-tree whose root is the column root ($R_A$ for \texttt{amount} in Figure~\ref{fig:merkletree}). Each column root is then hashed together with its column name, e.g., \texttt{"column:amount:"} followed by $R_A$, and the top-level tree is built over these column leaves. Binding each column root to its \emph{name}, not only its position, means that renaming, reordering, or swapping columns changes the top root even when the multiset of values is unchanged. Every data row must also have exactly as many fields as the header, so that no field (e.g., one appended after the last column) is left outside the tree; a file that violates this is hashed row by row instead, and for a file anchored column-wise the verifier reports such a row as tampering.

\item \emph{Structured, tensor-wise} is used for every \texttt{safetensors} model. A \texttt{safetensors} file starts with an 8-byte length field, followed by a JSON header that lists, for every tensor, its name, data type (\texttt{dtype}), \texttt{shape}, and byte range (\texttt{data\_offsets}) in the data section that follows. Figure~\ref{fig:tensortree} shows this layout for a small model with three tensors.
The header is parsed directly, without an external library. Each tensor becomes one leaf in two steps: first, the tensor's bytes are hashed; then this hash is combined with the tensor's name, data type, and shape, and hashed again. For example, for \texttt{attn.weight} the system hashes its 256~bytes, i.e., $8 \times 8 = 64$ values of 4~bytes each, shown as bytes [32--288] in Figure~\ref{fig:tensortree}, and then hashes the text \texttt{"tensor:attn.weight:F32:[8,8]:"} followed by that hash. Since the name, type, shape, and bytes all go into the leaf, changing any of them changes the leaf. The leaves are sorted by tensor name, and the tree is built on top of them.
\item \emph{Flat, row-based} is used for TXT and JavaScript Object Notation Lines (JSONL) datasets, and for a CSV/TSV file only as a fallback when it has no header, no data rows, or a row whose number of fields differs from the header. Each non-empty line of the file, including the header, is hashed as one leaf together with its line number; for example, the line \texttt{0,spam,84.1} on line~1 is hashed as \texttt{"row:1:0,spam,84.1"}. A change is therefore localized to a line but not to the field within it: changing \texttt{spam} reports only that line~1 changed.
\item \emph{Flat, chunk-based} is used for all other files, including model formats other than \texttt{safetensors} (e.g., Open Neural Network Exchange (ONNX), Hierarchical Data Format (HDF5), PyTorch). Each 1~MB chunk of the file is one leaf, so a change is localized only to a 1~MB region.
\end{itemize}

\subsection{Blockchain Anchoring}
\label{sec:contract}
The root is anchored by a purpose-built Solidity contract, \texttt{MerkleIntegrityRegistry}, 
whose core entry point is:

\noindent\begin{minipage}{\linewidth}
\begin{lstlisting}[basicstyle=\ttfamily\scriptsize,columns=fullflexible]
function storeRecord(
  bytes32 merkleRoot,
  string calldata fileName,
  FileType fileType,
  HashScheme hashScheme,
  string calldata contentId
) external returns (uint256 recordId);
\end{lstlisting}
\end{minipage}

Each call appends an immutable record, which includes root, file name, dataset-or-model \texttt{FileType}, \texttt{HashScheme}, optional IPFS \texttt{contentId}, owner address, and timestamp, to an append-only list. To find records without scanning the whole list, the contract also keeps two lookup tables: one from each Merkle root to the records that anchor it, and one from each owner address to the records that this owner has anchored. A verifier can thus ask, in a single read call, which records contain a given root, by the method \texttt{getRecordIdsByRoot}, or which records a given owner has anchored, by the method \texttt{getRecordIdsByOwner}. Transactions are signed in the owner's browser wallet. 
The contract has no privileged administrator, so authority comes from the owner address (Section~\ref{sec:threat}).

\subsection{Distributed Storage} 
\label{sec:ipfs}
Anchoring a 32-byte root establishes \emph{what} the correct content is, but not \emph{where} to obtain it. For this, the artifact is stored on a DSFS. Any DSFS that addresses files by their content could be used; we chose IPFS~\cite{benet2014ipfs} because it is the most widely adopted one, it is public and permissionless like the blockchain it complements, and a file's CID is itself a cryptographic commitment to its content, so a retrieved file can be checked against its identifier. When the owner hashes a file, the bytes are optionally added and pinned (CIDv1, raw codec) on a local IPFS node run with Kubo, the reference IPFS implementation (formerly \texttt{go-ipfs}), which the application starts and stops together with itself; the CID is then stored on-chain with the root. Because Kubo announces new content to the DHT only on a periodic schedule, which can lag on a fresh node, the backend also issues an immediate \texttt{routing/provide} call after each pin. A verifier can obtain the file from (i) its own local copy, (ii) the pinning node itself, independently of DHT propagation, or (iii) a public gateway (\texttt{dweb.link}). IPFS is optional; without it, the rest of the pipeline is unaffected.

\subsection{End-to-End Store and Verify Flow}
\label{sec:flow}
\begin{figure}[t]
\centering
\begin{tikzpicture}[
  ostep/.style={draw, rounded corners, fill=blue!7, align=center, font=\scriptsize, minimum width=2.6cm, minimum height=7mm, inner sep=2pt},
  vstep/.style={draw, rounded corners, fill=orange!10, align=center, font=\scriptsize, minimum width=2.6cm, minimum height=7mm, inner sep=2pt},
  arr/.style={-{Latex[length=1.8mm]}, thin},
  lbl/.style={font=\scriptsize\bfseries}
]
\node[lbl] (l1) at (0.05,0.55) {Owner};
\node[ostep] (s1) at (0,0) {1.\ Hash file\\(client-side)};
\node[ostep, below=4mm of s1] (s2) {2.\ Pin to IPFS\\$\rightarrow$ CID (optional)};
\node[ostep, below=4mm of s2] (s3) {3.\ storeRecord()\\on-chain};
\draw[arr] (s1)--(s2); \draw[arr] (s2)--(s3);

\node[lbl] (l2) at (2.8,0.15) {Verifier};
\node[vstep] (v1) at (5,0) {4.\ Read record\\(root, scheme, CID)};
\node[vstep, below=4mm of v1] (v2) {5.\ Fetch bytes\\(local / IPFS)};
\node[vstep, below=4mm of v2] (v3) {6.\ Recompute root\\(declared scheme)};
\node[vstep, below=4mm of v3] (v4) {7.\ Match?};
\node[vstep, below=4mm of v4, fill=red!10] (v5) {8.\ Localize:\\column/row/tensor};
\draw[arr] (v1)--(v2); \draw[arr] (v2)--(v3); \draw[arr] (v3)--(v4);
\draw[arr] (v4)--node[right,font=\tiny]{mismatch}(v5);
\draw[arr] (s3.east) -- ++(0.3,0) |- (v1.west);
\end{tikzpicture}
\caption{End-to-end store and verify flow. The verifier's scheme choice (step 6) is read from the chain, not guessed, so structured and flat records both remain independently verifiable.}
\label{fig:flow}
\end{figure}
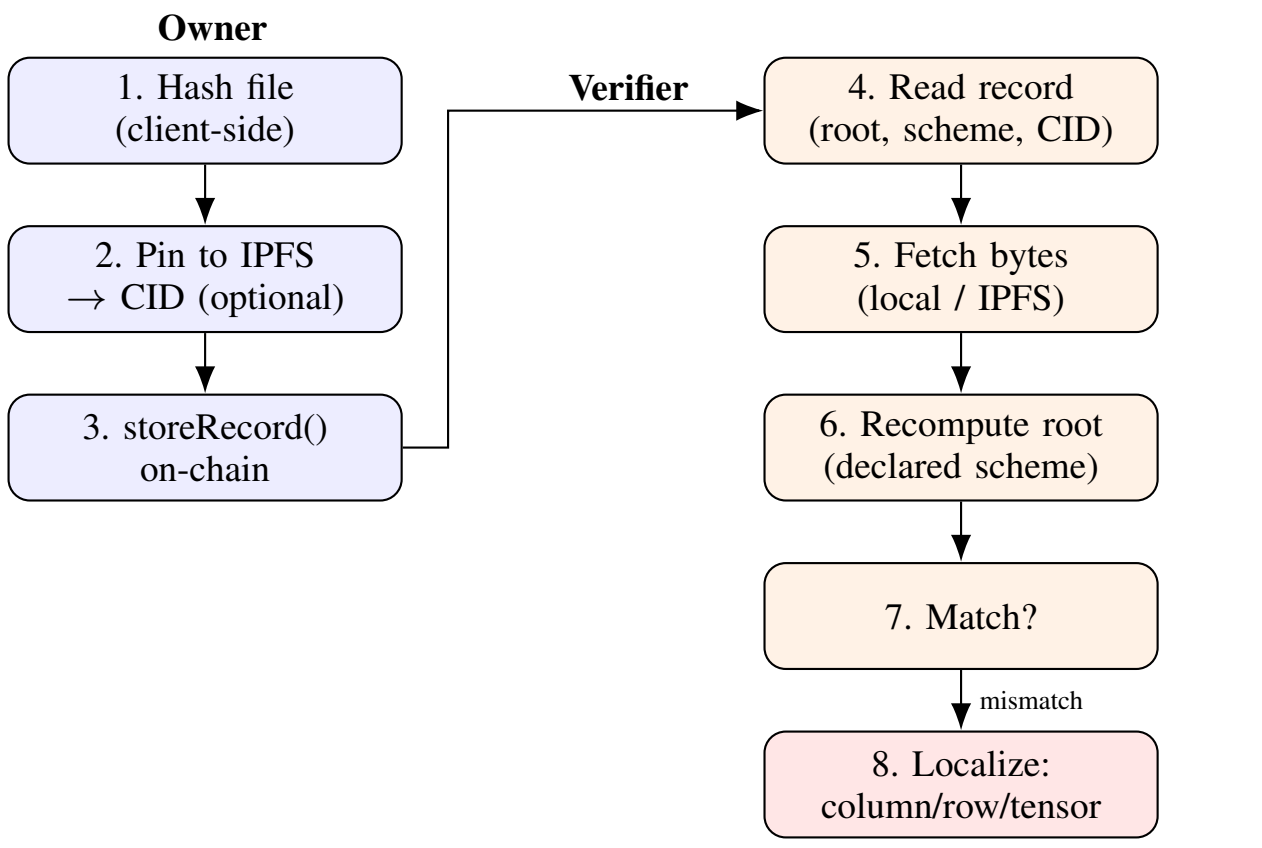

Figure~\ref{fig:flow} summarizes the two operations of the system. To store an artifact, the owner's client selects the scheme (Section~\ref{sec:hashing}) and computes the root locally. If IPFS is available, the file is pinned and a CID is obtained (Section~\ref{sec:ipfs}). The owner then signs a \texttt{storeRecord} transaction (Section~\ref{sec:contract}) that anchors the root, metadata, scheme, and CID.

To verify an artifact, the verifier selects a record and reads its metadata from the chain, including the \texttt{HashScheme} to use. The bytes are obtained from a local copy or the anchored CID, the root is recomputed with the declared scheme, and the result is compared with the on-chain value. On a mismatch, a scheme-specific routine compares each column root, tensor leaf, or row/chunk leaf with those of the authentic copy 
and reports every unit that disagrees, e.g., ``column \texttt{label}, row 412'' or ``tensor \texttt{attn.weight}'', as explained in the following.

\subsection{Reference for Localization}
\label{sec:manifest}
Since owner and verifier are logically separate, 
the verifier has only the record and the file. The root tells the verifier \emph{whether} a file changed, but not \emph{where}. To find the changed column or tensor, the verifier needs the original tree, which cannot be recovered from a tampered file. It therefore rebuilds the original tree from an authentic copy of the file: either a copy it has already verified, or the owner's copy fetched from IPFS through the CID stored on-chain. In both cases the copy is accepted only if its root equals the on-chain root, so no trusted party is needed. The verifier then compares the two trees column by column, or tensor by tensor; the entries that differ are the changed columns or tensors. If no authentic copy is available, the verifier still detects the change but cannot localize it.

In the prototype, localization therefore requires downloading the authentic copy; publishing the owner's exported tree file, with one 32-byte hash per column or tensor, would avoid this (Section~\ref{sec: conclusion}).

\section{Results}
\label{sec: results}

The complete system of Section~\ref{sec: proposedmethodology} is implemented to 
evaluate the hashing and localization code: we conduct a correctness test, three performance experiments, and a set of harder tampering scenarios. 
The owner and the verifier roles are exposed as two logical modes of one web application sharing one Node.js backend, so that the pipeline can be exercised under our control. 

The first is a \emph{correctness test}. The owner side (\texttt{app.js}) and the verifier side (\texttt{verifier.js}) contain separate implementations of the hashing schemes, which must agree exactly. The test hashes the same CSV file (3 columns $\times$ 25 rows) and the \texttt{safetensors} file, containing 2 tensors, with both and checks that every leaf, sub-root, and root is identical. It then modifies one value and runs the verifier's localization routine end to end, which correctly reported, e.g., \emph{column ``score'', row 3} and \emph{tensor ``layer0.weight''}. All checks are passed.

The second consists of three \emph{performance experiments} on synthetic files of increasing size. \textbf{Experiment~A} uses CSV datasets with 50 rows and 3, 10, or 30 columns. \textbf{Experiment~B} uses \texttt{safetensors} models with 5, 20, or 80 tensors of 2--50~KB each. \textbf{Experiment~C} uses CSV datasets with 10 columns and 50, 200, or 800 rows. In each experiment, the proposed scheme is compared with a baseline:

\begin{itemize}
\item \emph{Proposed}: the column-wise tree for CSV datasets and the tensor-wise tree for \texttt{safetensors} models (Section~\ref{sec:hashing}).
\item \emph{Baseline}: a conventional (flat) Merkle tree over the same file, with one leaf per line for CSV and one per 1~MB chunk for models, so it can point only to a changed line or chunk.
\end{itemize}

\subsection{Metrics}
The evaluation is carried out based on the following quantities:

\begin{itemize}
\item \emph{Build time (ms)}: wall-clock time to compute the complete tree and its root from the untampered file; the \emph{overhead} is the ratio of the proposed build time divided by the baseline's.
\item \emph{Localization accuracy}: the fraction of trials in which exactly the modified unit was reported, with no extra or missing units: the cell or tensor for the proposed scheme, the row for the row-based baseline.
\item \emph{Cells (or tensors) to inspect}: once a scheme has reported where the change is, how many cells (or tensors) a person must still check by hand to find it; lower is better. The proposed scheme points to one cell or tensor, so the value is 1. The row-based baseline points to a whole row, so every cell in that row must be checked, e.g., 30 cells for a 30-column dataset. The chunk-based baseline points to a 1~MB chunk, so every tensor stored in that chunk must be checked.
\end{itemize}

Each configuration of Experiments A and B received 8 tampering trials. Each trial applied exactly one modification to a fresh copy of the untampered file: for datasets, the suffix \texttt{\_X} was appended to one randomly chosen cell; for models, one randomly chosen byte of tensor data was incremented by one. The tampered file was re-hashed, and its tree was compared with the tree of the untampered file, which plays the role of the authentic copy (Section~\ref{sec:manifest}). Since each trial changes one unit and every unit has its own leaf, 100\% accuracy is expected by construction; the trials confirm that the implementation achieves it. Harder cases are examined separately in Section~\ref{sec:scenarios}.

All measurements use the unmodified production hashing code, run in a Node.js \texttt{vm} sandbox with a minimal browser shim. Files and tamper targets come from a seeded random generator (seed 1337), so every result is reproducible. Build times are single runs; the median over the 8 tampered trials agrees within 5\%, except for the smallest model (1.6 vs.\ 1.3~ms).

\subsection{Datasets: Localization and Overhead (Experiment A)}
Table~\ref{tab:experimentA} shows the results for CSV datasets. Both schemes found the modification in all 8 trials, but with different precision. The baseline reports only the changed \emph{row}, leaving as many cells to inspect as the number of columns, whereas the proposed tree reports the exact \emph{column and row}. The price is a longer build, in the range of 4.1$\times$ to 29.7$\times$, because every cell is hashed separately.

\begin{table}[h]
\centering
\caption{Experiment A: CSV datasets with 50 rows. Proposed = column-wise tree; Baseline = flat row-based tree.}
\label{tab:experimentA}
\small
\begin{tabular}{|l|c|c|c|}
\hline
\rowcolor{gray!50}
\textbf{Number of columns} & \textbf{3} & \textbf{10} & \textbf{30} \\\hline
Build time, proposed (ms) & 21.8 & 66.4 & 194.8 \\\hline
Build time, baseline (ms) & 5.3 & 6.7 & 6.6 \\\hline
Overhead (proposed / baseline) & 4.1$\times$ & 10.0$\times$ & 29.7$\times$ \\\hline
Localization accuracy, both & 100\% & 100\% & 100\% \\\hline
Cells to inspect, proposed & 1 & 1 & 1 \\\hline
Cells to inspect, baseline & 3 & 10 & 30 \\\hline
\end{tabular}
\end{table}

\subsection{Models: Localization and Overhead (Experiment B)}
Table~\ref{tab:experimentB} shows the results for \texttt{safetensors} models. The proposed tree identified the modified tensor in all 8 trials. The 5- and 20-tensor models are smaller than 1~MB, so the whole file is one chunk and every tensor remains a candidate. The 1.94~MB model spans two chunks, so on average 40.5 of its 80 tensors remain. The overhead is lower than for datasets (3.3--5.7$\times$), because each tensor is hashed as one contiguous byte range.

\begin{table}[h]
\centering
\caption{Experiment B: \texttt{safetensors} models. Proposed = tensor-wise tree; Baseline = flat 1~MB chunk-based tree. $^{*}$Average / maximum over 8 trials.}
\label{tab:experimentB}
\small
\setlength{\tabcolsep}{3pt}
\begin{tabular}{|l|c|c|c|}
\hline
\rowcolor{gray!50}
\textbf{Number of tensors} & \textbf{5} & \textbf{20} & \textbf{80} \\\hline
File size & 140 KB & 526 KB & 1.94 MB \\\hline
Build time, proposed (ms) & 1.6 & 4.7 & 16.5 \\\hline
Build time, baseline (ms) & 0.5 & 1.2 & 2.9 \\\hline
Overhead (proposed / baseline) & 3.3$\times$ & 4.1$\times$ & 5.7$\times$ \\\hline
Localization accuracy, proposed & 100\% & 100\% & 100\% \\\hline
Tensors to inspect, proposed & 1 & 1 & 1 \\\hline
Tensors to inspect, baseline$^{*}$ & 5 / 5 & 20 / 20 & 40.5 / 43 \\\hline
\end{tabular}
\end{table}

\subsection{Scalability with Dataset Size (Experiment C)}
For a 10-column dataset with 50, 200, and 800 rows, the proposed tree took 67, 248, and 995~ms to build, versus 7, 25, and 98~ms for the baseline. Both grow linearly with the row count, and the proposed tree stays about 10$\times$ slower (10 cells hashed per line); even 800 rows took under one second.


\subsection{Harder Tampering Scenarios}
\label{sec:scenarios}
Beyond single substitutions, we applied the eight harder edits of Table~\ref{tab:scenarios} with the same code and fixed inputs. All were detected, and their localization falls into three groups. First, edits that keep the file's structure (several cells or tensors at once, or a tensor's dtype) were localized exactly. Second, renamed or swapped units were attributed to the right unit, but not to a single cell: every cell leaf contains its column name, so renaming a column changes all its cells and all 20 rows are reported; swapping two columns reports both in full, and a renamed tensor is reported under its new name. The verifier thus shows \emph{where} the change is, but not that it is a rename or a swap. Third, every cell leaf also contains its row index, so inserting or deleting a row shifts and flags every later row, not only the edited one.

\begin{table}[h]
\centering
\caption{Harder tampering scenarios (CSV: 20 rows, 4 columns; model: 5 tensors). All edits were detected; the right column shows what localization reported. For row insertion and deletion, counts vary by column because some shifted values equal those they replace.}
\label{tab:scenarios}
\small
\setlength{\tabcolsep}{3pt}
\begin{tabular}{|l|l|}
\hline
\rowcolor{gray!50}
\textbf{Edit} & \textbf{Localization result} \\\hline
3 cells in 3 columns & exact: the 3 cells \\\hline
1 row inserted & all later rows (12--16 of 20) \\\hline
1 row deleted & all later rows (11--15 of 20) \\\hline
2 columns swapped & both columns, all 20 rows \\\hline
1 column renamed & that column, all 20 rows \\\hline
2 tensors modified & exact: the 2 tensors \\\hline
1 tensor F32$\to$F16 & exact: that tensor \\\hline
1 tensor renamed & that tensor (new name) \\\hline
\end{tabular}
\end{table}

\subsection{Discussion and Limitations}
\label{sec:discussion}
A signed flat list of per-column or per-tensor hashes would localize single-unit changes equally well; the Merkle tree's advantage, one 32-byte root whatever the number of units, is modest at our scales, and we did not benchmark this baseline.

The signature and provenance tools answer a different question. Sigstore~\cite{newman2022sigstore}, its proposed use for ML models~\cite{meiklejohn2025mlsupplychain}, and in-toto~\cite{torresarias2019intoto}, record \emph{who} published an artifact and \emph{how} it was produced. To do so, the publisher signs a digest of the artifact, usually a hash of the whole file, and the signature is recorded in a public transparency log~\cite{laurie2013rfc6962}. These tools provide the identity and key management, but a tampered file still appears only as a digest mismatch. The two approaches are therefore complementary: the structure-aware root can be the digest that the publisher signs, giving both a verified publisher and the location of any change.

Overall, a one-time, sub-second owner overhead buys exact localization at every hand-off. The evaluation has clear limitations: the column-wise overhead grows with column count; only CSV/TSV and \texttt{safetensors} are structured; artifacts were under 2~MB; an inserted or deleted row marks every later row as changed; and anchoring ran only on a testnet, with both roles in one deployment. Finally, both building blocks are well established; our contribution is their combination for ML artifacts, with name-bound leaves and an on-chain scheme tag.

\section{Conclusion and Future Directions}
\label{sec: conclusion}
This paper presented a blockchain-anchored, structure-aware Merkle tree that lets an independent verifier not only detect that a dataset or ML checkpoint changed but also localize the change to a named column, row, or tensor. On the production code, cell and tensor changes were localized exactly at a one-time build overhead of 3--30$\times$ over a flat Merkle tree, and inserted or deleted rows were detected but marked every later row as changed.
Future work follows four directions. The first is \emph{breadth and scale}: extending structure-aware hashing to other widely used formats, namely Apache Parquet, a column-oriented format for tabular data; ONNX, a common format for model graphs and weights; HDF5, used, e.g., by Keras; and GPT-Generated Unified Format (GGUF), the format of quantized large language models; and parallelizing hashing with WebAssembly~\cite{haas2017webassembly}, a portable binary code format that browsers run at near-native speed, to keep gigabyte-scale artifacts practical. The second is \emph{deployment}: a field trial with owner and verifier in separate organizations, publishing the exported tree file on IPFS, and measuring mainnet gas costs. The third is \emph{trust}: binding each root to a publisher identity, e.g., through Sigstore, with key management, rotation, and revocation, and covering a malicious owner and a compromised blockchain or broken hash function, e.g., by anchoring on several chains and switching hash functions through the scheme tag. The fourth is \emph{richer localization}: classifying each divergence (e.g., added, renamed, or re-quantized), so that a deployment pipeline can automatically block a serious change, such as modified weights, and only warn about a harmless one, such as a renamed column; aligning rows by content so that inserted or deleted rows are pinpointed; and evaluating the whole pipeline against real poisoning and backdoor attacks.

\section*{Acknowledgment}
This work is partially supported by the European Commission through the HORIZON-MSCA-SE EXARCH project (GA: 101236244).

\bibliography{bibliography}

@misc{nakamoto2008bitcoin,
  title={Bitcoin: A Peer-to-Peer Electronic Cash System},
  author={Nakamoto, Satoshi},
  year={2008},
  howpublished={\url{https://bitcoin.org/bitcoin.pdf}}
}

@misc{buterin2014ethereum,
  title={Ethereum: A Next-Generation Smart Contract and Decentralized Application Platform},
  author={Buterin, Vitalik},
  year={2014},
  howpublished={Ethereum Whitepaper}
}

@inproceedings{merkle1988digital,
  title={A Digital Signature Based on a Conventional Encryption Function},
  author={Merkle, Ralph C.},
  booktitle={Advances in Cryptology --- CRYPTO '87},
  series={Lecture Notes in Computer Science},
  volume={293},
  pages={369--378},
  year={1988},
  publisher={Springer}
}

@article{benet2014ipfs,
  title={IPFS -- Content Addressed, Versioned, P2P File System},
  author={Benet, Juan},
  journal={arXiv preprint arXiv:1407.3561},
  year={2014}
}

@article{wang2024blockchain,
  title={Blockchain-Based Dynamic Cloud Data Integrity Auditing via Non-Leaf Node Sampling of Rank-Based Merkle Hash Tree},
  author={Wang, Chenxu and others},
  journal={IEEE Transactions on Network Science and Engineering},
  year={2024}
}

@inproceedings{han2024blockchain,
  title={Blockchain-Based Data Integrity Verification},
  author={Han, Zhijie and others},
  booktitle={2024 4th International Conference on Blockchain Technology and Information Security (ICBCTIS)},
  year={2024}
}

@article{liu2023data,
  title={Data Integrity Audit Scheme Based on Quad Merkle Tree and Blockchain},
  author={Liu, Zhenpeng and others},
  journal={IEEE Access},
  year={2023}
}

@article{mohan2020merkle,
  title={Merkle Tree and Blockchain-Based Cloud Data Auditing},
  author={Mohan, A. and others},
  journal={International Journal of Cloud Applications and Computing},
  year={2020}
}

@article{wei2020blockchain,
  title={Blockchain Data-Based Cloud Data Integrity Protection Mechanism},
  author={Wei, Pengcheng and others},
  journal={Future Generation Computer Systems},
  year={2020},
  publisher={Elsevier}
}

@article{he2020blockchain,
  title={Blockchain Based Data Integrity Verification for Cloud Storage with T-Merkle Tree},
  author={He, Kai and others},
  journal={Prototype Implementation and Evaluation},
  year={2020}
}

@article{kumar2020blockchain,
  title={Blockchain-Based Framework for Data Storage in Peer-to-Peer Scheme Using Interplanetary File System},
  author={Kumar, Randhir and others},
  journal={Handbook of Research on Blockchain Technology},
  year={2020}
}

@inproceedings{nyaletey2019blockipfs,
  title={BlockIPFS -- Blockchain-Enabled Interplanetary File System for Forensic and Trusted Data Traceability},
  author={Nyaletey, Emmanuel and others},
  booktitle={2019 IEEE International Conference on Blockchain (Blockchain)},
  year={2019},
  organization={IEEE}
}

@article{saif2024efficient,
  title={Efficient and Secure Distributed Data Storage and Retrieval Using Interplanetary File System and Blockchain},
  author={Saif, Muhammad Bin and others},
  journal={Future Internet},
  year={2024}
}

@article{zhu2026research,
  title={Research on an On-Chain and Off-Chain Collaborative Storage Method Based on Blockchain and IPFS},
  author={Zhu, Tianqi and others},
  journal={Future Internet},
  year={2026}
}

@article{meiklejohn2025mlsupplychain,
  title={Machine Learning Models Have a Supply Chain Problem},
  author={Meiklejohn, Sarah and others},
  journal={arXiv preprint arXiv:2505},
  year={2025}
}

@article{hanailscg2026,
  title={AI Supply Chain Galaxy: 3D Visual Analytics for License Compliance},
  author={Han, Wei and others},
  journal={arXiv preprint},
  year={2026}
}

@article{goldblum2020dataset,
  title={Dataset Security for Machine Learning: Data Poisoning, Backdoor Attacks, and Defenses},
  author={Goldblum, Micah and others},
  journal={IEEE Transactions on Pattern Analysis and Machine Intelligence},
  year={2020}
}

@article{cina2022wild,
  title={Wild Patterns Reloaded: A Survey of Machine Learning Security Against Training Data Poisoning},
  author={Cin{\`a}, Antonio Emanuele and others},
  journal={ACM Computing Surveys},
  year={2022}
}

@article{tian2022comprehensive,
  title={A Comprehensive Survey on Poisoning Attacks and Countermeasures in Machine Learning},
  author={Tian, Zhiyi and others},
  journal={ACM Computing Surveys},
  year={2022}
}

@article{yerlikaya2022data,
  title={Data Poisoning Attacks Against Machine Learning Algorithms},
  author={Yerlikaya, Fahri An{\i}l and Bahtiyar, Serif},
  journal={Expert Systems with Applications},
  year={2022},
  publisher={Elsevier}
}

@article{kermani2015systematic,
  title={Systematic Poisoning Attacks on and Defenses for Machine Learning in Healthcare},
  author={Mozaffari Kermani, Mehran and others},
  journal={IEEE Journal of Biomedical and Health Informatics},
  year={2015}
}

@inproceedings{newman2022sigstore,
  title={Sigstore: Software Signing for Everybody},
  author={Newman, Zachary and Meyers, John Speed and Torres-Arias, Santiago},
  booktitle={Proceedings of the 2022 ACM SIGSAC Conference on Computer and Communications Security (CCS)},
  pages={2353--2367},
  year={2022}
}

@inproceedings{torresarias2019intoto,
  title={in-toto: Providing Farm-to-Table Guarantees for Bits and Bytes},
  author={Torres-Arias, Santiago and Afzali, Hammad and Kuppusamy, Trishank Karthik and Curtmola, Reza and Cappos, Justin},
  booktitle={Proceedings of the 28th USENIX Security Symposium},
  pages={1393--1410},
  year={2019}
}

@misc{laurie2013rfc6962,
  title={Certificate Transparency},
  author={Laurie, Ben and Langley, Adam and Kasper, Emilia},
  howpublished={RFC 6962, Internet Engineering Task Force},
  year={2013}
}

@inproceedings{haas2017webassembly,
  title={Bringing the Web up to Speed with {WebAssembly}},
  author={Haas, Andreas and Rossberg, Andreas and Schuff, Derek L. and Titzer, Ben L. and Holman, Michael and Gohman, Dan and Wagner, Luke and Zakai, Alon and Bastien, J. F.},
  booktitle={Proceedings of the 38th ACM SIGPLAN Conference on Programming Language Design and Implementation (PLDI)},
  pages={185--200},
  year={2017}
}

\end{document}